\documentclass[12pt]{article}
\usepackage{amssymb,amsmath,bm,epsfig,comment}

\renewcommand{\thefootnote}{\fnsymbol{footnote}}
\newcommand{\im}{{\rm Im\,}}
\newcommand{\re}{{\rm Re\,}}
\newcommand{\tr}{{\rm Tr\,}}
\begin{document}
\title{}

\title{
\begin{flushright}
\begin{minipage}{0.2\linewidth}
\normalsize
WU-HEP-15-15 \\*[50pt]
\end{minipage}
\end{flushright}
{\Large \bf 
Nonuniversal gaugino masses  in a magnetized toroidal \\
compactification of SYM theories 
\\*[20pt] } }

\author{
Keigo~Sumita\footnote{
E-mail address: k.sumita@aoni.waseda.jp}\\*[20pt]
{\it \normalsize 
Department of Physics, Waseda University, 
Tokyo 169-8555, Japan} \\*[50pt]}

\date{
\centerline{\small \bf Abstract}
\begin{minipage}{0.9\linewidth}
\medskip 
\medskip 
\small 
This paper proposes a concrete model of nonuniversal gaugino masses 
on the basis of higher-dimensional supersymmetric Yang-Mills theories 
compactified on a magnetized factorizable torus, 
and we estimate the gauge coupling constants and gaugino masses in the model. 
In the magnetized toroidal compactifications, 
the four-dimensional effective action can be obtained analytically identifying 
its dependence on moduli fields, where the magnetic fluxes are able to yield 
the flavor structure of the minimal supersymmetric standard model (MSSM). 
The obtained gauge kinetic functions contains multi moduli fields 
and their dependence is nonuniversal for the three gauge fields. 
The nonuniversal gauge kinetic functions can lead to 
nonuniversal gaugino masses at a certain high energy scale (e.g. compactification scale). 
Our numerical analysis of them shows that, 
particular ratios of gaugino masses, which were found to enhance the Higgs boson mass and 
lead to ``natural supersymmetry'' in the MSSM, can be realized in our model, 
while the gauge couplings are unified as is achieved in the MSSM. 
\end{minipage}}

\begin{titlepage}
\maketitle
\thispagestyle{empty}
\clearpage
\tableofcontents
\thispagestyle{empty}
\end{titlepage}

\renewcommand{\thefootnote}{\arabic{footnote}}
\setcounter{footnote}{0} 

\section{Introduction} 
Supersymmetry (SUSY) has been regarded as one of the strong candidates 
for new physics beyond the standard model (SM), 
and the minimal supersymmetric standard model (MSSM) is quite supported 
by many people. 
Indeed, SUSY models has been proposed enormously so far 
and many of them respect main structure of the MSSM. 

One of the great implications of SUSY is a cancellation of quantum corrections 
proportional to the square of cutoff scale. 
Low-energy SUSY ensures the stability of scalar masses, 
and the notorious fine-tunning problem 
for the stability of Higgs boson mass known in the SM 
is expected to be solved in the MSSM. 
From this point of view, we see that the low-energy SUSY is desirable. 
However, the mass of Higgs boson particle discovered 
at the Large Hadron Collider \cite{Aad:2012tfa,Chatrchyan:2012ufa} 
got the MSSM into a serious situation. 
The Higgs boson mass calculated at the tree-level of the MSSM is bounded by 
the $Z$ boson mass. A large amount of quantum corrections to the Higgs boson mass 
is required to realize the observed value. 
Dominant contributions of the quantum corrections is due to propagation of the SUSY particles. 
The Higgs boson mass is quite related to SUSY spectra in the MSSM or MSSM-like models. 
The sufficient amount of corrections requires a relatively heavy SUSY spectrum, but 
this is usually accompanied with another fine-tunning problem of the $\mu$-parameter 
(a SUSY higgsino mass parameter), that is, the so-called little hierarchy problem. 
As a result, 
a fine-tunning which is much harder than $\mathcal O(1\%)$ cannot be avoided now 
in most of conventional parameter space of the MSSM.

Another of the great impacts of SUSY in the MSSM is the gauge coupling unification. 
SUSY requires the presence of SUSY partners of the SM particles, and 
in the presence of such light particles charged under the SM gauge groups 
the renormalization group (RG) flows for the gauge coupling constants become 
different from those in the SM. 
As a result, it is well know that 
the three gauge couplings can be unified at a certain high energy scale 
within low-scale SUSY breaking scenarios. 
The unifying scale is about $M_{\rm GUT} = 2.0\times 10^{16}{\rm GeV}$ in the MSSM.

In the unifications given by SUSY, we usually assume that the three gaugino masses are 
also unified at the same or near that scale. This popular assumption seems to be sensible, 
indeed, this has been adopted in many SUSY phenomenological studies. 
However, there is no reason for them to be constrained to unify, and 
we can freely choose their values at the $M_{\rm GUT}$ scale 
as input parameters in the MSSM. 
Considering a recent circumstances of particle physics experiments, 
that is, no detection of the SUSY particles and the relatively heavy mass of the Higgs boson, 
probably we are also required to investigate unconventional parameter space of the MSSM, 
i.e., nonuniversal gaugino masses.

The generic framework of nonuniversal gaugino masses were studied 
so far in the MSSM \cite{Abe:2007kf,Abe:2012xm,Antusch:2012gv} and 
these works found an attractive feature. 
The gaugino masses are related to the Higgs boson mass in RG flows, 
and a certain range of the mass ratios will enhance the Higgs boson mass 
while a typical mass scale of SUSY spectra is not so much high. 
As the result, they found the fine tunning of $\mu$-parameter discussed above 
would be relaxed very well.

We consider unified theories behind the SM or ultraviolet (UV) complete theories 
in particle physics. 
In the top-down approaches of phenomenological studies on the basis of such theories, 
particle physics models are obtained as four-dimensional (4D) low-energy effective action. 
In such model building, the gaugino masses as well as the other masses and couplings 
cannot be set arbitrary, 
which should be determined by other structure of the theories, e.g., extra dimensional space. 
Although it is attractive that the natural parameter regions are still alive in the MSSM, 
we must remark how to realize such rations of gaugino masses 
as a boundary condition at the $M_{\rm GUT}$ scale 
on the basis of such theories for a high energy physics. 
This paper provides a realization of nonuniversal gaugino masses 
on the basis of higher-dimensional SYM theories 
which appear in low-energy limits of superstring theories. 

In model building based on higher-dimensional SYM theories, 
structure of extra dimensions of space is the most significant issue 
to obtain a realistic model with 4D chiral spectra like the SM. 
It is known that toroidal compactifications with magnetic fluxes 
are able to yield such a spectrum in higher-dimensional SYM theories 
\cite{Bachas:1995ik,Cremades:2004wa}. 
The magnetized toroidal compactification have been actively studied, 
and some concrete models have been constructed, 
where the flavor structure of the SM, such as, the three generations of matters and 
their hierarchical spectrum, are obtained. 
In particular, Ref.~\cite{Abe:2012fj,Abe:2014soa} proposed such a model 
based on a 10D $U(8)$ SYM theories, 
and the mass spectrum of the SUSY particles as well as the SM particles 
were studied to verify this model. 
These works owe mainly their model building to 
an $\mathcal N=1$ superfield description of ten-dimensional (10D) 
magnetized SYM theories given in Ref.~\cite{Abe:2012ya}. 
That proposed a systematic way of dimensional reduction of the magnetized SYM theories 
with the superfield description, 
and derived the 4D effective action identifying its dependence on 
a dilaton and moduli superfields. 
In the superspace formulation, 
an $\mathcal N=1$ SUSY out of the full $\mathcal N=4$ SUSY 
(counted by the 4D supercharges) of 10D SYM theories is described manifestly, 
and $\mathcal N=1$ SUSY configurations of magnetic fluxes is facilitated 
to study.

The superfield description for 10D SYM theories have been recently extended 
to be able to apply to (4+2n)-dimensional SYM theories and their mixtures 
which are well motivated by D-brane pictures of superstring theories\cite{Abe:2015jqa}. 
This extension allows a large variety of model building, and especially, 
we find that the mixture is a suitable foundation to construct a model of 
nonuniversal gaugino masses. This paper shows 
such a concrete model based on a mixture of a six-dimensional (6D) SYM theory 
and a 10D SYM theory compactified on magnetized tori, 
and estimates the three SM gauge coupling constants and 
the gaugino masses.

This paper is constructed as follows. 
We briefly review the nonuniversal gaugino masses in the MSSM in Sec.~2. 
Sec.~3 gives an overview of higher-dimensional SYM theories 
compactified on a factorizable torus with magnetic fluxes. 
This section shows how to realize the flavor structure of the SM by 
magnetic fluxes. 
In Sec.4, the superfield description of the magnetized SYM theories 
is briefly reviewed. The way of dimensional reduction and identifying 
the moduli dependence will become clear. 
Sec.~5 is devoted to show our results. 
A concrete model will be proposed, 
where we study the gaugino masses to verify that our model 
accommodate a realization of the nonuniversal gaugino masses 
desired from the phenomenological point of view. 
Sec.~6 contains conclusions and discussions.

\section{Nonuniversal gaugino masses in the MSSM} 
We shortly review nonuniversal gaugino masses in the MSSM, 
which shows one of our motivations of this work. 
In a popular assumption, three gaugino masses are set to be unified 
at the $M_{\rm GUT}$ scale for simplicity, 
\begin{equation*}
M_1(M_{\rm GUT})=M_2(M_{\rm GUT})=M_3(M_{\rm GUT}), 
\end{equation*}
where $M_i$ ($i=1,2,3$) are the gaugino mass parameters of $U(1)_{Y}$, 
$SU(2)_{L}$ and $SU(3)_{C}$, respectively. 
However, they are just free parameters and nothing restricts their values 
to be degenerate in the MSSM. 
It is also possible to consider nonuniversal gaugino masses in the MSSM.

The nonuniversal gaugino masses in the MSSM 
were studied in Refs.~\cite{Abe:2007kf,Abe:2012xm,Antusch:2012gv}, 
where they found that 
nonuniversal gaugino masses with certain ratios enhance 
the Higgs boson mass through the RG effects, 
and the fine-tunning of so-called $\mu$-parameter will be relaxed. 
To discuss more details, we define the degree of fine-tunning as 
$100/\Delta_\mu(\%)$\cite{Barbieri:1987fn} with 
\begin{equation*}
\Delta_\mu=\left|\frac{\partial\log m_{\rm Z}^2}{\partial\log\mu^2}\right|. 
\end{equation*}
In Ref.~\cite{Abe:2012xm}, the authors studied impact of the nonuniversal gaugino masses 
on the Higgs boson mass and the SUSY spectrum in the light of the LHC data 
with a simple ansatz for other parameters, e.g., Yukawa couplings. 
As the result, in the range of 
\begin{equation}
3.0 \lesssim \frac{M_2(M_{\rm GUT})}{M_3(M_{\rm GUT})} \lesssim 5.5,\qquad 
-3.0 \lesssim \frac{M_1(M_{\rm GUT})}{M_3(M_{\rm GUT})},
\label{eq:ratio1}
\end{equation}
the $126$ GeV Higgs boson will be realized with the tunning of $\mathcal O(1)(\%)$. 
Furthermore, in the following region, 
\begin{equation*}
5.2 \lesssim \frac{M_2(M_{\rm GUT})}{M_3(M_{\rm GUT})} \lesssim 5.5, 
\end{equation*}
the fine-tunning of $\mu$-parameter will be relaxed as well as $\mathcal O(10)(\%)$.

In the generic framework of SUSY models, 
we require the presence of hidden sector sequestered from the MSSM sector, 
where SUSY should be spontaneously broken. 
The SUSY breaking contribution is mediated to the MSSM sector, and thus, 
the mediation mechanism determines the SUSY spectrum 
as a boundary condition of RG flows at a certain high energy scale. 

Three mediation mechanisms due to moduli fields, 
conformal anomaly~\cite{Randall:1998uk,Giudice:1998xp} and 
gauge interactions\cite{Dine:1981gu} 
are famous and available in generic frameworks. 
These mediation mechanisms and their combinations have been actively studied so far, 
and we know their phenomenological features precisely. 
According to that, 
it is impossible to realize the the desirable ratio $M_2/M_3 \sim 5$ 
within the anomaly and gauge mediation mechanisms, 
and we see that the SUSY breaking contributions mediated by moduli fields 
through nonuniversal gauge kinetic functions are indispensable to obtain 
the ideal ones.

When the gauge kinetic functions of $SU(3)_{C}$, $SU(2)_{L}$ and $U(1)_{Y}$ 
can be flexibly chosen differently from each other 
in terms of its dependence on moduli fields, the gaugino masses are controllable. 
However, in models based on the unified theories, 
the couplings of moduli fields to the MSSM fields are determined 
by structure of extra dimensional space. 
We give a realization of nonuniversal gaugino masses 
on the basis of a concrete model derived from a higher-dimensional SYM system 
compactified on magnetized tori.

\section{Higher-dimensional SYM theories on magnetized tori} 
This section reviews higher-dimensional SYM theories compactified on 
two-dimensional (2D) tori with magnetic fluxes. 
Higher-dimensional fields are expanded into zero-modes and multiple Kaluza-Klein (KK) modes. 
In this section, we derive 4D effective actions from the magnetized SYM theories 
focusing on the zero-modes. KK modes are considered to be heavy enough 
to be decoupled in the low-energy effective field theories. 
To show typical features of magnetized toroidal compactifications, 
we concentrate on a two-dimensional torus whose coordinates are denoted by $(x,y)$. 
The line element is given by 
\begin{equation*}
ds^2=g_{ij}dx^idx^j,
\end{equation*}
where the metric $g$ is 
\begin{equation*}
g=(2\pi R)^2\begin{pmatrix}
1&\re\tau\\
\re\tau&|\tau|^2
\end{pmatrix}. 
\end{equation*}
Parameters $\tau$ and $R$ determine the shape and size of this torus. 

We consider a 2D spinor field on this torus,
\begin{equation*}
\psi=\begin{pmatrix}
\psi_+\\
\psi_-
\end{pmatrix}.
\end{equation*}
The zero-mode equations for these fields are given by 
the Dirac operator on this internal space as 
\begin{eqnarray}
\bar\partial_z\psi_++[\bar A_z, \psi_+]&=&0\label{eq:zeroii}\\
\partial_z\psi_--[A_z, \psi_-]&=&0,\label{eq:zeroij}
\end{eqnarray}
where we use complex coordinates $z$ and 
2D complex vector $A_z$ defined by 
two real coordinates $x,y$ of this torus and 
two components of vector fields $(A_x,A_y)$ as follows, 
\begin{eqnarray*}
z&\equiv&\frac12\left(x+\tau y\right)\\
A_z&\equiv&-\frac1{\im\tau}\left(\tau^*A_x-A_y\right). 
\end{eqnarray*}
Toroidal periodicity for the two coordinates is expressed 
by $z\sim z+1$ and $z\sim z+\tau$. 

Considering a nontrivial configuration of gauge potential, 
we introduce magnetic fluxes on this torus, 
and the configuration is 
\begin{equation*}
A_z=\frac{\pi}{\im\tau}\left(M\bar z+\bar \zeta\right), 
\end{equation*}
where the magnetic flux is given by an $N\times N$ matrix $M$ 
in $U(N)$ theories. 
A continuous Wilson lines can also be introduced, 
which is identified as a constant term of gauge potential and 
denoted by an $N\times N$ matrix $\zeta$ here. 
This paper considers only simple Abelian forms of the flux and Wilson line 
because those are sufficient and necessary to construct particle physics models, 
thus offdiagonal entries of $M$ and $\zeta$ are set to be vanishing. 
Note that the nonvanishing entries of $M$ must be integer values because of 
the Dirac's quantization condition. 
In the case of $N=2$, if the two diagonal elements of flux matrix take different values from 
each other, $U(2)$ gauge symmetry is broken down to $U(1)\times U(1)$. 
Thus, in generic $U(N)$ theories, 
the magnetic fluxes $M$ can lead to gauge symmetry breaking as 
$U(N)\rightarrow \prod_a U(N_a)$ 
($U(N_a)$ is a remaining gauge subgroup of $U(N)$). 
This Wilson lines $\zeta$ are also able to induce such a gauge symmetry breaking 
in the same way as the magnetic fluxes.

In the zero-mode equation (\ref{eq:zeroii}) for $\psi_+$, 
we can elicit a bifundamental representation $(N_a, \bar N_b)$ 
of $U(N_a)\times U(N_b)$ as 
\begin{eqnarray}
[\bar\partial_z+\frac{\pi }{2\im\tau}\left(M_{ab}z + \zeta_{ab}\right)] 
\left(\psi_+\right)_{ab}&=&0, 
\end{eqnarray}
where a bifundamental representation $(N_a, \bar N_b)$ contained in $\psi_+$ is 
denoted by $\left(\psi_+\right)_{ab}$. 
A magnetic flux felt by it 
is defined as $M_{ab}\equiv M_{a}-M_{b}$, 
and a Wilson line is also defined as $\zeta_{ab}\equiv \zeta_{a}-\zeta_{b}$. 
According to Ref.~\cite{Cremades:2004wa}, this equation has $M_{ab}$ 
normalizable solutions when $M_{ab}>0$, and then, 
zero-modes of the same representation $(N_a,\bar N_b)$ contained in the other spinor $\psi_-$ 
are eliminated 
because a relative sign of the zero-mode equation (\ref{eq:zeroij}) is different. 
On the contrary, when $M_{ab}<0$, 
$\left(\psi_-\right)_{ab}$ has $|M_{ab}|$ well-defined zero-modes and 
the other $\left(\psi_+\right)_{ab}$ has none. Thus, 
the magnetic fluxes cause a kind of chirality projection which is to generate 
a 4D chiral spectrum like the SM. 

The degenerate zero-modes appear corresponding to the magnitude of fluxes, 
which we can identify with the generations of the SM. 
Their wavefunctions can be obtained analytically, 
expressed by using the Jacobi-theta functions. 
Moreover, they have a Gaussian profile on the torus, and 
their localized points on the magnetized torus are different from each other. 
Since overlap integrals of the zero-mode wavefunctions determine the magnitude of 
their 4D effective couplings, 
zero-modes localized far away from each other yield a suppressed coupling, 
which can give a hierarchical structure to Yukawa couplings. 
It is also attractive that 
the integrals of zero-mode wavefunctions on the magnetized torus 
can be performed analytically. 
The simple expression for Yukawa couplings \cite{Cremades:2004wa} and 
higher-order couplings \cite{Abe:2009dr} were obtained. 
Summarizing the above, the magnetic fluxes are able to yield the flavor structure of the SM, 
such as, the three generations and their hierarchical masses and mixing angles. 

\section{Superfield description of SYM systems}
This section introduces a superfield description of higher-dimensional SYM theories 
on the magnetized tori\cite{Abe:2012ya}, 
which is extremely useful, especially for constructing a particle physics model 
with an $\mathcal N=1$ SUSY vacuum configuration. 
A 4D effective action with $\mathcal N=1$ SUSY 
can be derived from magnetized SYM theories in the superfield description, 
identifying its dependence on the moduli fields.

Higher-dimensional field theories have higher-dimensional SUSY as 
$\mathcal N=2,3$ and 4 counted by the 4D supercharges, but 
these theories can be described in the 4D $\mathcal N=1$ superspace formulation 
focusing on an $\mathcal N=1$ SUSY of the whole higher-dimensional SUSY 
in the following way. First, we consider a 10D SYM action compactified on three 2D tori, 
\begin{equation}
S = \int d^{10}X\sqrt {-G} \left\{
-\frac1{4g^2}\tr\left(F^{MN}F_{MN}\right)
+\frac{i}{2g^2}\left(\bar\lambda\Gamma^MD_M\lambda\right)
\right\}, \label{eq:10dsym}
\end{equation}
where $g$ is the gauge coupling constant. 
Capital Latins $M,N$ run for the 10D spacetime coordinate, and 
the 10D field strength $F^{MN}$, covariant derivative $D_M$ and 
gamma matrix $\Gamma^M$ are contracted by the 10D metric $G_{MN}$. 
The 10D line element is 
\begin{equation*}
ds^2 = \eta_{\mu\nu}dx^\mu dx^\nu + g_{mn}dx^mdx^n, 
\end{equation*}
where $\mu, \nu : 0,1,2,3$ and $m,n : 4,5,\ldots,9$. 
The 4D Minkowski spacetime is given by $\eta = {\rm diag} (-,+,+,+)$ and 
$g_{mn}$ gives three tori, $T^2\times T^2\times T^2$. 

Field contents of this theory is given by 
a 10D vector field $A_M$ and 
a Majorana-Weyl spinor field $\lambda$, 
satisfying conditions $\Gamma^{10}\lambda=+\lambda$ and 
$\lambda^{C}=\lambda$ ($\Gamma^{10}$ is a 10D chirality operator and 
C represents the charge conjugate. ). 
To make an $\mathcal N=1$ SUSY manifest, 
the 10D vector field is decomposed into a 4D vector field and three complex fields as 
\begin{equation*}
A_\mu,\qquad A_i\equiv-\frac1{\im\tau_i}\left(\tau_i^*A_m-A_n\right), 
\end{equation*}
where $i=1,2,3$ and $(m,n)=(2+2i,3+2i)$. 
The 10D spinor field is also decomposed into 
the following four 4D Weyl spinor fields, 
\begin{equation*}
\lambda_0\equiv\lambda_{+++},\qquad
\lambda_1\equiv\lambda_{+--},\qquad
\lambda_2\equiv\lambda_{-+-},\qquad
\lambda_3\equiv\lambda_{--+},
\end{equation*}
where signs $\pm$ represent a chirality on each torus, 
e.g., $\lambda_{+--}$ has a positive chirality on the first torus and negative ones 
on the other tori. 4D Spinor fields with the other chirality are not contained 
in the 10D Majorana-Weyl spinor because of $\Gamma^{10}\lambda=+\lambda$. 

The component fields form 4D $\mathcal N=1$ vector multiplets 
and chiral multiplets, 
\begin{equation*}
\left\{A\mu,\lambda_0\right\},\qquad
\left\{A_i,\lambda_i\right\}. 
\end{equation*}
These supermultiplets are embedded into a vector superfield and 
three chiral superfields as follows, 
\begin{eqnarray*}
V&\equiv& -\theta\sigma^\mu\bar\theta A_\mu +i\bar\theta\bar\theta\theta\lambda_0
-i\theta\theta\bar\theta\bar\lambda_0+\frac12\theta\theta\bar\theta\bar\theta D,\\
\phi_i&\equiv& \frac1{\sqrt2} A_i +\sqrt2\theta\lambda_i +\theta\theta F_i, 
\end{eqnarray*}
where two-component spinors $\theta$ and $\bar\theta$ are 
4D $\mathcal N=1$ supercoordinates. 

The 10D SYM action (\ref{eq:10dsym}) is expressed in the 4D $\mathcal N=1$ 
superspace formulation with the above superfields as 
\cite{Marcus:1983wb,ArkaniHamed:2001tb}, 
\begin{equation}
S=\int d^{10}X \sqrt {-G}
\left[ \int d^4\theta \mathcal K 
+\left\{ \int d^2\theta\left(\frac1{4g^2}\mathcal W^\alpha\mathcal W_\alpha
+\mathcal W\right) +{\rm h.c.}\right\}\right],
\end{equation}
where three functions $\mathcal K$, $\mathcal W$ and $\mathcal W^\alpha$ 
are given by 
\begin{eqnarray}
\mathcal K &=& \frac2{g^2}h^{\bar i j}\tr\left[
\left(\sqrt2\bar\partial_{\bar i}+\bar\phi_{\bar i}\right)e^{-V}
\left(-\sqrt2\partial_j+\phi_j\right)e^V
+\bar\partial_{\bar i}e^{-V}\partial_j e^V \right]+\mathcal K_{\rm WZW}, \nonumber\\
\mathcal W &=& \frac1{g^2}\epsilon^{\rm ijk}e_{\rm i}^{~i}e_{\rm j}^{~j}
e_{\rm k}^{~k} \tr\left[\sqrt2\phi_i\left(\partial_j\phi_k-\frac1{3\sqrt2}
\left[\phi_j, \phi_k\right]\right)\right],\nonumber\\
\mathcal W_\alpha &=& -\frac14 \bar D\bar De^{-V}D_\alpha e^V.\label{eq:10dsym2}
\end{eqnarray}
In these expressions, $h^{\bar ij}$ and $e_{\rm i}^{~i}$ express 
the metric and fielbein of each torus. 
The supercovariant derivatives are denoted by 
$D_\alpha$ and $\bar D_{\dot\alpha}$. 

The field equations for auxiliary fields $D$ and $F_i$ are 
given by 
\begin{eqnarray*}
D=-h^{\bar i j}\left(\bar\partial_{\bar i}A_j+\partial_j\bar A_{\bar i}
+\frac12[\bar A_{\bar i}, A_j]\right),\\
\bar F_{\bar i} = -h_{j\bar i}\epsilon^{\rm jkl}e_{\rm j}^{~j}e_{\rm k}^{~k}e_{L}^{~l} 
\left(\partial_kA_l-\frac14[A_k,A_l]\right). 
\end{eqnarray*}
The $\mathcal N=1$ SUSY expressed by the superfield formulation is preserved 
as long as the Vacuum Expectation Values (VEVs) of 
these auxiliary fields are vanishing. 

The magnetic fluxes are introduced as 
\begin{equation}
\langle A_i\rangle=\frac{\pi}{\im\tau_i}\left(M^{(i)}\bar z_i + \bar\zeta^{(i)}\right). \label{eq:aivev}
\end{equation}
These magnetic fluxes should satisfy a condition $\langle D\rangle=\langle F_i\rangle=0$ 
to preserve the $\mathcal N=1$ SUSY. 
On this magnetized background, 
the zero-mode equations for the superfield $\phi_j$ 
on the $i$-th torus are obtained as 
\begin{eqnarray}
\left[\bar\partial_{\bar i} +\frac{\pi}{2\im\tau_i}
\left(M_{ab}^{(i)}z_i + \zeta^{(i)}_{ab}\right)\right](f_j^{(i)})_{ab} &=& 0 \qquad{\rm for}\quad i=j,
\label{eq:zerosuperii}\\
\left[\partial_i -\frac{\pi}{2\im\tau_i}
\left(M_{ab}^{(i)}\bar z_{\bar i} + \bar\zeta^{(i)}_{ab}\right) \right]
(f_j^{(i)})_{ab} &=& 0 \qquad{\rm for}\quad i\neq j, \label{eq:zerosuperij}
\end{eqnarray}
where $(f_j^{(i)})_{ab}$ represents a zero-mode wavefunction on the $i$-th torus 
of a bifundamental representation $(N_a, \bar N_b)$ contained in $\phi_j$. 
These equations for $i=j$ and $i\neq j$ 
have the same form as the Dirac equations (\ref{eq:zeroii}) and (\ref{eq:zeroij}), 
respectively. 
Thus, their zero-mode wavefunctions are obtained 
and the 4D effective action 
can also be derived analytically in this superfield description. 
The gauge fields of the MSSM are accommodated by diagonal parts of 
the vector superfield $V$ which do not feel the Abelian magnetic fluxes. 
Offdiagonal component of $V$ get heavy because of the partial gauge symmetry breaking, 
and they will be decoupled. 

The 4D effective action generically contains some parameters. such as, 
the gauge coupling constant, torus radius and complex structure. 
These parameters are promoted to moduli fields in supergravity framework. 
It is known that a conventional relation between the VEVs of moduli fields 
and the parameters as follows\cite{Ibanez:2012zz}, 
\begin{equation*}
\re \langle S\rangle =e^{-\langle\phi\rangle}\alpha'^{-3}\mathcal A^{(i)},\qquad
\re \langle T_r\rangle =e^{-\langle\phi\rangle}\alpha'^{-1}\mathcal A^{(r)},\qquad
\re \langle U_r\rangle =i\bar\tau_r,
\end{equation*}
where $\mathcal A^{(r)}$ represents the area of the $r$-th torus and 
the 10D dilaton field determines the 10D gauge coupling, 
\begin{equation*}
g=e^{\langle\phi\rangle/2}\alpha'^{3/2}. 
\end{equation*}
This relation allows us to determine the moduli dependence of 
the 4D effective actions. 
Thus, explicit forms of gauge kinetic functions, K\"ahler metrics and holomorphic 
Yukawa couplings as functions of the moduli fields are obtained.

This systematic procedure to derive 4D effective supergravity 
actions from 10D SYM theories compactified on magnetized tori was extended 
to apply to $(4+2n)$-dimensional SYM theories and their mixtures in Ref.~\cite{Abe:2015jqa}. 
In that paper, two specific 
4D effective supergravity actions with the explicit moduli dependence 
were shown to demonstrate the extended procedure. 
They were derived from two SYM systems well motivated by 
stable D-brane systems. 
One is a mixture of a 4D SYM theory and an eight-dimensional SYM theory. 
The other is composed of a 6D SYM theory and a 10D SYM theory. 
In this paper, we consider the latter SYM system to realize 
the nonuniversal gaugino masses (Note the two systems are essentially equivalent 
because they should be related to each other by T-duality). 

\section{A concrete model} 
We construct an MSSM-like model based on a mixture of a 10D SYM theory compactified on 
three tori $(T^2)_1\times (T^2)_2\times (T^2)_3$ and a 6D SYM theory on $(T^2)_1$. 
Although the D-brane physics is one of our motivations to consider this SYM system, 
we are just starting from the SYM theories here 
and will not mention whole consistency for string models in this paper. 

\subsection{Pati-Salam models}
One of the most important things in such model buildings 
in the magnetized toroidal compactifications is that all the flavor structure must originate 
from a single torus. 
For example, if the three generations of left-handed quarks are 
generated by magnetic fluxes on the first torus and those of right-handed quarks 
are induced on the second torus, rank of their Yukawa matrix is reduced to one, 
and two of the three generations will remain massless 
after the electroweak symmetry breaking. 
Thus, to construct a three-generation model in the SYM system given in the top of this section, 
the magnetic fluxes on the first torus $(T^2)_1$ 
must yield whole the flavor structure of the MSSM, 
and the magnetic fluxes on the other tori should be determined not to 
disturb the structure and to satisfy conditions to preserve the $\mathcal N=1$ SUSY.

A unique configuration of magnetic fluxes to generate 
whole the flavor structure of SM on a single torus 
was found~\cite{Abe:2012ya}\footnote{
Orbifold projections lead to other flux configurations to 
construct three-generation models \cite{Abe:2008fi,Abe:2008sx,Abe:2015yva}. } 
and an MSSM-like model was proposed starting from a 10D $U(8)$ SYM theory. 
The $U(8)$ gauge group was broken down to the Pati-Salam gauge group, 
$U(4)_{C}\times U(2)_{L}\times U(2)_{R}$, which is to lead to a SM-like gauge 
group $SU(3)_{C}\times SU(2)_{L}\times (U(1))^5$ by introducing Wilson lines. 
To give an overview of the model, we define 
the following matrix $M^{(i)}$ which appears in the VEV of Eq.~(\ref{eq:aivev}), 
\begin{equation*}
M^{(i)}=\begin{pmatrix}
m^{(i)}_{C}\times{\bm 1}_4 &0&0\\
0&m^{(i)}_{L}\times{\bm 1}_2&0\\
0&0&m^{(i)}_{R}\times{\bm 1}_2  
\end{pmatrix}.
\end{equation*}
The suitable configuration is then shown as 
\begin{eqnarray}
\left(m^{(1)}_{C},m^{(1)}_{L},m^{(1)}_{R}\right) &=& \left(0,+3,-3\right)\nonumber
\end{eqnarray}
In the Pati-Salam group, 
$(4,\bar 2,1)$ representation contains the left-handed matter fields and 
$(\bar 4,1,2)$ representation does the right-handed matters. 
Higgs multiplets are carried by $(1,2,\bar 2)$ representation, 
and each representations feel the magnetic fluxes of 
$m_{C}^{(i)}-m_{L}^{(i)}$, 
$m_{R}^{(i)}-m_{C}^{(i)}$ and 
$m_{L}^{(i)}-m_{R}^{(i)}$, respectively. 
We summarize them in Table \ref{tb:10dmag}. 
\begin{table}[htb]
\center
\begin{tabular}{cccc} \hline
Representations & MSSM fields & Fluxes on $(T^2)_1$&\# of gen.\\\hline 
$(4,\bar 2,1)$ & Left-handed & $m_{C}^{(1)}-m_{L}^{(1)}=-3$&3  \\
$(\bar 4,1,2)$ & Right-handed & $m_{R}^{(1)}-m_{C}^{(1)}=-3$&3 \\
$(1,2,\bar 2)$ & Higgs & $m_{L}^{(1)}-m_{R}^{(1)}=+6$&6\\\hline
  \end{tabular}
\caption{Field contents and magnetic fluxes felt by them on the first torus 
in the Pati-Salam model.}\label{tb:10dmag}
\end{table}
On the first torus, these magnetic fluxes induce three-generations of 
left- and right-handed matters and six-generations of Higgs fields. 
The presence of these multiple Higgs fields is a generic feature of the magnetized SYM theory 
as well as D-brane models, and we identify a linear-combinations of the six ones 
with the MSSM Higgs multiplet. 

In the single 10D SYM model, flux configurations on the other two tori 
were determined as 
\begin{eqnarray}
\left(m^{(2)}_{C},m^{(2)}_{L},m^{(2)}_{R}\right) &=& \left(0,-1,0\right)\nonumber\\
\left(m^{(3)}_{C},m^{(3)}_{L},m^{(3)}_{R}\right) &=& \left(0,0,+1\right),
\end{eqnarray}
which will not spoil the flavor structure generated on the first torus 
and satisfy the SUSY preserving condition $\langle D\rangle = \langle F_i\rangle =0$ 
with the following relation of the torus area, 
\begin{equation*}
\mathcal A^{(1)}/\mathcal A^{(2)} = \mathcal A^{(1)}/\mathcal A^{(3)} =3.
\end{equation*}

We respect the Pati-Salam gauge and this configuration of magnetic fluxes also in this paper, 
and those can be embedded into a mixture of 10D $U(N)$ SYM theory compactified on 
three tori and 6D $U(8-N)$ theory on $(T^2)_1$. 
Thus, we first divide the Pati-Salam group into two parts. 
Table \ref{tb:10dmag} reads that 
the Higgs multiplets contained in representation $(1,2,\bar 2)$ feel 
positive magnetic fluxes, $+6$, on the first torus. 
This means that the Higgs fields originate from $\phi_1$ which is 
a superfield consisting of $A_1$ and $\lambda_1$, 
because the other chiral superfields require negative magnetic fluxes 
to have zero-modes on the first torus. 
In the action of 10D $U(N)$ and 6D $U(8-N)$ SYM theories described in the 
superspace formulation, 
all bifundamental representations of $U(N)\times U(8-N)$ are completely 
eliminated in $\phi_1$. 
This is because the bifundamental representations are to form 
a hypermultiplet under the 6D (4D $\mathcal N=2$) SUSY but 
the chiral superfield $\phi_1$ is consists of vector components of the 6D theory. 
Therefore, the Higgs multiplets, $(1,2,\bar 2)$ representations of the Pati-Salam gauge group, 
must be contained in adjoint representations of $U(N)$ or $U(8-N)$. 
We can determine from this discussion that 
the Pati-Salam model is embedded into the SYM system, 
being divided into $U(4)_{C}$ and $U(4)_{LR}\supset U(2)_{L}\times U(2)_{R}$.

With this division of the Pati-Salam gauge group, 
two types of SYM systems are available : 
One consists of 
the 10D $U(4)_{C}$ SYM and the 6D $U(4)_{LR}$ SYM theories, 
and the other of the 10D $U(4)_{LR}$ SYM and the 6D $U(4)_{C}$ SYM theories. 
In the former case, $U(4)_{LR}$ has to be broken down 
to $U(2)_{L}\times U(2)_{R}$ by the magnetic fluxes on the first torus, 
which fluxes must also break the $\mathcal N=1$ SUSY 
because there is no other contribution of magnetic fluxes in the $U(4)_{LR}$ sector 
to satisfy $\langle D\rangle=0$. 
On the contrary, magnetic fluxes of the $U(4)_{LR}$ sector are given on three tori 
in the other case. 
Therefore, only the latter case can take SUSY preserving configurations. 
We have found the suitable SYM system, which is studied in the next subsection. 

\subsection{A model of nonuniversal gaugino masses}
We consider the SYM system consisting of 
the 10D $U(4)_{LR}$ SYM and the 6D $U(4)_{C}$ SYM theories, which 
are compactified on magnetized $(T^2)_1\times (T^2)_2\times (T^2)_3$ and 
$(T^2)_1$, respectively. 
The configuration of magnetic fluxes is parametrized as 
\begin{equation*}
M^{(1)}_{C} = m_{C}\times{\bm 1}_4, 
\end{equation*}
and 
\begin{equation*}
M^{(i)}_{LR}=\begin{pmatrix}
m^{(i)}_{L}\times{\bm 1}_2&0\\
0&m^{(i)}_{R}\times{\bm 1}_2 
\end{pmatrix}, 
\end{equation*}
where $M^{(1)}_{C}$ is the magnetic flux in the $U(4)_{C}$ SYM theory, 
and $M^{(i)}_{LR}$ ($i=1,2,3$) are those in $U(4)_{LR}$ SYM theory, 
on each torus. 
These parameters are set as 
\begin{eqnarray*}
m_{C} &=& 0\\
\left(m^{(1)}_{L},m^{(1)}_{R}\right) &=& \left(+3,-3\right)\\
\left(m^{(2)}_{L},m^{(2)}_{R}\right) &=& \left(-1,0\right)\\
\left(m^{(3)}_{L},m^{(3)}_{R}\right) &=& \left(0,+1\right). 
\end{eqnarray*}
The magnetic fluxes on the first torus $M^{(1)}_{C}$ and $M^{(1)}_{LR}$ 
yield the same flavors as is shown in Table~\ref{tb:10dmag}.  
The others are determined to preserve the $\mathcal N=1$ SUSY 
without changes of the flavor structure. 
The vanishing $m_{C}$ is also a key to preserve the SUSY.

Although there remain a few massless adjoint fields other than the MSSM fields 
in the 4D effective theory, 
all chiral exotic fields are eliminated. 
In the model derived from the 10D $U(8)$ SYM theory, 
both of them appeared, and this is an advantage of our model. 
The remaining massless adjoint fields are called open-string moduli. 
The presence of them is well known as a notorious problem in string models. 
Although we might propose some prescription to eliminate them, e.g., 
orbifold projections\cite{Abe:2008fi}, we just assume that they vanish somehow in this paper.

Now, we are able to derive the 4D effective supergravity action, 
and its spectrum can be calculated exactly. 
The K\"ahler metrics and holomorphic Yukawa couplings 
obtained in this model 
are almost the same as the 10D $U(8)$ SYM model 
except for some trivial numerical factors. 
These were shown in Ref.~\cite{Abe:2012fj,Abe:2014soa}, 
and one can calculate the explicit form in accordance with Ref.~\cite{Abe:2015jqa}. 
The phenomenological features of the 10D $U(8)$ SYM model were studied precisely, 
and a consistent spectrum of the SM particles and their partners was 
obtained\cite{Abe:2012fj,Abe:2014soa}. 
However, the gauge kinetic functions are very different. 
The model proposed in this section leads to nonuniversal gauge kinetic functions, 
while they were universal in the 10D $U(8)$ SYM model. 
The gauge kinetic functions of 
$SU(3)_{C}$ and $SU(2)_{L}$ are straightforwardly obtained as 
\begin{eqnarray}
f_3&=&T_1,\nonumber\\
f_2&=&S.\label{eq:gaugekine23}
\end{eqnarray}

Before showing that of $U(1)_{Y}$ gauge group, 
we should explain breaking of $U(4)_C$ and $U(2)_L\times U(2)_R$. 
We can introduce the Wilson line parameters to break the Pati-Salam gauge group 
on the first torus. 
As the result, $U(4)_{C}$ is broken down to $U(3)_{C}\times U(1)_{C'}$, 
and $U(2)_{R}$ gauge is also broken to $U(1)_{R'}\times U(1)_{R''}$. 
Thus, there are five $U(1)$ gauge symmetries : 
\begin{eqnarray*}
U(1)_a&\equiv& U(1)\subset U(3)_{C},\qquad
U(1)_b\equiv U(1)_{C'},\qquad
U(1)_c\equiv U(1)\subset U(2)_{L},\\
U(1)_d&\equiv& U(1)_{R'},\qquad
U(1)_e\equiv U(1)_{R''}.
\end{eqnarray*}
We can define the hypercharge $Q_Y$ by a linear combination of these five 
$U(1)$ symmetries as\footnote{
The other four linear combinations are expected to be decoupled from the low-energy 
effective theory (see Section~3.3 of \cite{Abe:2012fj})} 
\begin{equation}
Q_Y = x Q_a +\left(x-\frac23\right)Q_b +\left(x-\frac16\right)Q_c 
+\left(x-\frac23\right)Q_d+\left(x+\frac13\right)Q_e,\label{eq:linear}
\end{equation}
where $x$ is an arbitrary number. 
The charges for each content are summarized in Table~\ref{tb:charge}. 
\begin{table}[htb]
\center
\renewcommand{\arraystretch}{1.2}
\begin{tabular}{ccccccc} \hline
Contents & $Q_a$ & $Q_b$ & $Q_c$ & $Q_d$ & $Q_e$ & $Q_Y$\\\hline 
$Q$ & 1 & 0 & -1 & 0 & 0 & $1/6$  \\
$L$ & 0 & 1 & -1 & 0 & 0 & -$1/2$  \\
$H_u$ & 0 & 0 & 1 & -1 & 0 & $1/2$  \\
$H_d$ & 0 & 0 & 1 & 0 & -1 & -$1/2$  \\
$u$ & -1 & 0 & 0 & 1 & 0 & -$2/3$  \\
$d$ & -1 & 0 & 0 & 0 & 1 & $1/3$  \\
$\nu$ & 0 & -1 & 0 & 1 & 0 & $0$  \\
$e$ & 0 & -1 & 0 & 0 & 1 & $1$  \\\hline
 \end{tabular}
\renewcommand{\arraystretch}{1}
\caption{The charges of the five original $U(1)$ symmetries and $U(1)_{Y}$ hypercharges 
defined in Eq.~(\ref{eq:linear}) are shown. 
}\label{tb:charge}
\end{table}
Its gauge coupling constant $g_Y$ is then given by 
\begin{equation}
\frac{1}{g_Y^2} = \frac{x^2}{{\tilde g_{\rm6D}}^2} + \frac{(x-2/3)^2}{{\tilde g_{\rm6D}}^2} 
+\frac{(x-1/6)^2}{{\tilde g_{\rm10D}}^2}+\frac{(x-2/3)^2}{{\tilde g_{\rm10D}}^2}
+\frac{(x+1/3)^2}{{\tilde g_{\rm10D}}^2}, \label{eq:hypergg}
\end{equation}
where $\tilde g_{\rm 6D}$ and $\tilde g_{\rm10D}$ are 
are 4D effective gauge coupling constants of the 6D $U(4)_{C}$ 
and the 10D $U(4)_{LR}$ SYM theories given by 
\begin{equation*}
\tilde g_{\rm6D} =e^{\langle\phi\rangle/2}\sqrt{\frac{\alpha'}{\mathcal A^{(1)}}},\qquad
\tilde g_{\rm10D} =e^{\langle\phi\rangle/2}\sqrt{\frac{{\alpha'}^3}{\mathcal A^{(1)}\mathcal A^{(2)}
\mathcal A^{(3)}}}. 
\end{equation*}
The arbitrary number $x$ used in Eq.~(\ref{eq:linear}) never appear in the hypercharges 
but it effects only on the gauge coupling, that is, we can control it with this parameter. 
From Eq.~(\ref{eq:hypergg}), the $U(1)_{Y}$ gauge kinetic function reads 
\begin{equation}
f_1=(2x^2-\frac43x+\frac49)T_1+(3x^2-\frac43x+\frac7{12})S \label{eq:gaugekine1}, 
\end{equation}
where the two coefficient given as functions of parameter $x$ should be 
rational numbers to be consistent with the discrete axionic shift symmetries. 
We will discuss this constraint again later. 

In the rest of this subsection 
we discuss some typical features of spectrum determined 
by the K\"ahler metrics and holomorphic Yukawa couplings. 
As we have mentioned 
it is known that a consistent spectrum is obtained with them, 
but changes of RG flows due to nonuniversal gauge kinetic functions 
might affect on the spectrum. 
Moreover, it is nontrivial that the gaugino masses can be varying 
independently of the other soft parameters at the $M_{\rm GUT}$ scale.

RG flows of the Yukawa couplings may be deflected compared with the MSSM, 
with the nonuniversal gauge kinetic functions. 
Its deflection cannot be drastic because 
the running of the gauge couplings are hardly changed as will show later. 
A semi-realistic spectrum of the quarks and leptons 
obtained in Ref.~\cite{Abe:2012fj} can also be realized in our model.

SUSY spectra of this model are given in the moduli mediation or a combination 
with the anomaly mediation which is called 
mirage mediation scenarios\cite{Choi:2005uz,Endo:2005uy}. 
RG effects on the softparameters will be changed in the case of 
nonuniversal gaugino masses. 
Although discussions for SUSY spectra in the 10D $U(8)$ model 
are unable to apply directly to our model, 
it seems that the deviation will never lead to dangerous tachyons and flavor violations, 
unless a bizarre boundary condition for squared scalar mass and 
A-terms are realized in our model. 
In the following, we remark on possibilities to realize such a boundary condition 
by the mixed contributions of $S$ and $T_1$. 

As we will see, SUSY breaking contributions mediated by the moduli fields $S$ and $T_1$ 
must be determined to have a certain ratio in order to 
obtain desirable nonuniversal gaugino masses. 
In accordance with Ref.~\cite{Abe:2014soa}, dilaton superfield $S$ properly 
contributes to all the softparameters, and 
the CMSSM like-spectrum is realized by the sole contribution of $S$. 
On the contrast, 
K\"ahler moduli field $T_1$ induces 
negative contributions to some of squared softmasses, 
and this cannot dominate SUSY breaking contributions in order to forbid the presence of 
tachyonic modes. 
Fortunately, the favored ratio of gaugino masses ($M_2>M_3$) 
and gauge kinetic functions~(\ref{eq:gaugekine23})
imply that 
the contribution of dilaton $S$ is much larger than that of K\"ahler moduli $T_1$. 

Even when the $T_1$ contribution is subdominant, 
it might lead to unfamiliar boundary conditions of 
the soft parameters (other than the gaugino masses) at a compactification scale, 
which has a possibility to spoil the great advantage 
of non-universal gaugino masses shown in Sec.~2. 
Although we can estimate the exact spectrum, 
it is not necessary because 
it is easy to remove $T_1$ SUSY breaking contributions to 
all the squared scalar masses and A-terms 
while the moduli still generates the gaugino masses. 
There are three K\"ahler moduli fields, $T_1$, $T_2$ and $T_3$ in our model, 
and we consider SUSY breaking contributions mediated by them. 
When  they are equal to each other, that is, 
\begin{equation*}
\frac{F^{T_1}}{t_1+\bar t_1} = \frac{F^{T_2}}{t_2+\bar t_2} = 
\frac{F^{T_3}}{t_3+\bar t_3}, 
\end{equation*} 
where $F^{T_i}$ and $t_i$ are the VEVs of the auxiliary field 
and the lowest component of moduli superfield $T_i$, 
their contributions are completely canceled\cite{Abe:2014soa}. 
The modulus $T_1$ generates the gaugino masses, 
while its SUSY breaking contribution is canceled out by 
the other contributions of $T_2$ and $T_3$ in the soft scalar masses and the A-terms. 
Thus, we can control nonuniversal gaugino masses independently of 
the boundary conditions of the other soft parameters.

\subsection{Gauge coupling unification and gaugino masses}
This section analyzes the nonuniversal gauge kinetic functions. 
In studies of nonuniversal gaugino masses 
\cite{Abe:2007kf,Abe:2012xm,Antusch:2012gv}, 
the gauge coupling constants are set to be unified as is in the MSSM. 
The exact gauge coupling unification is not essentially important but 
drastic changes of the gauge couplings may not relax the fine-tunning of 
the $\mu$-parameter. 
At least, without the gauge coupling unification, 
the specific ratios of the gaugino masses (\ref{eq:ratio1}) 
would not be reliable. 
To discuss the gauge coupling unification, 
we define the compactification scale in this model by the mass scale of the lightest KK mode as, 
\begin{equation*}
M_C = 1/\sqrt {\mathcal A^{(1)}}. 
\end{equation*}
The magnetic fluxes and the Wilson lines have the mass scale as high as $M_C$. 
Thus, the gauge groups are broken down at this scale.

The easiest way to realize the gauge coupling unification in our model is that 
two types of 4D effective gauge coupling constants, $\tilde g_6$ and $\tilde g_{10}$, 
are chosen to a unified value which 
the MSSM predicts at the $M_{\rm GUT}$ scale, $4\pi/g_a^2\sim24$, 
and the SUSY breaking and the compactification scales are set to be $10^3$ GeV and 
$M_{\rm GUT}$, respectively. (Note that, the SUSY breaking scale here is defined as 
a typical scale at which the SUSY particles are decoupled from the SM particles.) 

In the D-brane picture of type IIB string theory, 
more constraints will be imposed on these scales and 
gauge coupling constants. We are studying the SYM theories but 
it is worth verifying the consistency. 
In the framework of type IIB string theory, the 4D Planck mass scale and 
the 4D effective gauge coupling of Dp-branes are given by (Note that $g_5$ and 
$g_9$ given in this expression are equivalent to 
the 4D effective couplings of 6D and 10D SYM theories, 
$\tilde g_6$ and $\tilde g_{10}$, respectively.)
\begin{eqnarray}
M_{\rm pl} &=&\frac{2e^{-2\phi}}{(2\pi)^7{\alpha'}^4}V,\label{eq:mplanck}\\
\frac1{g_p^2} &=& \frac{e^{-\phi}}{(2\pi)^{p-2}{\alpha'}^{(p-3)/2}}V_{p-3}, 
\label{eq:gauge}
\end{eqnarray}
where $e^\phi=g_{\rm st}$ and $\alpha'=1/M^2_{\rm st}$, and 
$g_{\rm st}$ and $M_{\rm st}$ represent a string coupling and a string scale, respectively. 
$V$ is the volume of 6D extra compact space, and
$V_{p-3}$ denotes the volume of cycles which the Dp-branes wrap. 
In these expressions, 
when we set the two gauge couplings as $4\pi^2/g^2_5=4\pi^2/g^2_9=24$, 
taking the realistic value of 4D Planck mass and the SUSY condition 
$\mathcal A^{(1)}/\mathcal A^{(2)}=\mathcal A^{(1)}/\mathcal A^{(3)}=3$ into account, 
we find 
\begin{eqnarray}
M_{\rm st} &\sim& 4.31\times 10^{18} {\rm GeV},\label{eq:mst} \\
M_C &\sim& 3.96\times 10^{17} {\rm GeV}. \label{eq:mcc}
\end{eqnarray}
The compactification scale is slightly deviated from the $M_{\rm GUT}$ scale. 
The gauge couplings evolve to the compactification scale $M_C$ 
through the $M_{\rm GUT}$ scale within the MSSM RG equations. 
Although this means that the gauge coupling unification might be missed, 
the exact form of gauge coupling unification is not required and 
it is sufficient to at least assert that the favored ratios of gaugino masses (\ref{eq:ratio1}) 
are also reliable in our model. 
Another estimation of these scales given in the Appendix supports this discussion.

We calculate the gaugino masses at the compactification scale. 
In the framework of supergravity, 
the gaugino masses are easily calculated by using a formula 
with the specific forms of 
gauge kinetic functions (\ref{eq:gaugekine23}) and (\ref{eq:gaugekine1}) 
as follows, 
\begin{equation*}
M_a=F^m\partial_m\ln\left(\re f_a\right) + \frac{b_ag_a^2}{16\pi^2}\frac{F^C}{C_0}. 
\end{equation*}
In the assumption of the gauge coupling unification like the MSSM, 
the gauge couplings $g_a$ and coefficients $b_a$ are given as 
$4\pi/g_a^2 = 24$ and $(b_1,\,b_2,\,b_3) = (33/5,\,1,\,-3)$, 
and we find 
\begin{eqnarray*}
M_3 &=&\frac{F^{T_1}}{t_1+\bar t_1} +\frac{-1}{32\pi}\frac{F^C}{C_0},\\
M_2 &=&\frac{F^{S}}{s+\bar s} +\frac{1}{96\pi}\frac{F^C}{C_0}\\
M_1 &=&(2x^2-\frac43x+\frac49)\frac{F^{T_1}}{t_1+\bar t_1} +(3x^2-\frac43x+\frac7{12})
\frac{F^{S}}{s+\bar s} 
+\frac{11}{160\pi}\frac{F^C}{C_0}, 
\end{eqnarray*}
where $s$ and $ t_1$ represents the VEVs of 
the lowest component fields of the moduli superfields $S$ and $T_1$, 
and $F^{S}$ and $F^{T_1}$ are the VEVs of their auxiliary fields. 
$C_0$ and $F^C$ are VEVs of the lowest component field and an auxiliary filed of 
a chiral compensator superfield of the supergravity. 
The following reparametrization simplifies the above expressions of 
gaugino masses, 
\begin{equation*}
M_{\rm SUSY} \equiv\frac{F^{T_1}}{t_1+\bar t_1} , \qquad 
R^S \equiv \frac1{M_{\rm SUSY}}\frac{F^{S}}{s+\bar s}, \qquad 
R^C \equiv \frac1{4\pi^2}\frac1{M_{\rm SUSY}}\frac{F^{C}}{C_0}. 
\end{equation*}
The gaugino masses are then expressed by
\begin{eqnarray*}
M_3 &=&M_{\rm SUSY}\left(1-\frac{\pi}{8}R^C\right),\\
M_2 &=&M_{\rm SUSY}\left(R^S+\frac{\pi}{24}R^C\right),\\
M_1 &=&M_{\rm SUSY}\left( (2x^2-\frac43x+\frac49) 
+(3x^2-\frac43x+\frac7{12})R^S+\frac{11\pi}{40}R^C\right). 
\end{eqnarray*}
Their ratios which is of our interest are independent of $M_{\rm SUSY}$. 
It should be determined to be of $\mathcal O(1\,{\rm TeV})$ being consistent 
with experimental results. 

We determine the value of parameter $x$ in the assumption of 
gauge coupling unification as follows. 
The gauge coupling unification $\langle\re f_a\rangle=6/\pi$ is 
realized by $\langle \re S\rangle=\langle \re T_1\rangle=6/\pi$ 
in (\ref{eq:gaugekine23}) and (\ref{eq:gaugekine1}), 
so we assume that these moduli fields are stabilized satisfying this relation 
by a stabilization mechanism somehow. 
In this assumption, we find the following condition on the parameter $x$ 
\begin{equation}
(2x^2-\frac43x+\frac49)+(3x^2-\frac43x+\frac7{12})=1,\label{eq:normal}
\end{equation}
which is equivalent to $x=(8\pm\sqrt{59})/30$. 
Note that, the value of $x$ should be a rational number 
for the coefficient of $U(1)_Y$ gauge kinetic functions ($2x^2-\frac43x+\frac49$ and 
$3x^2-\frac43x+\frac7{12}$ ) to be rational numbers, 
otherwise the discrete axionic shift symmetries of 
the moduli fields cannot be explained. 
However, any rational number cannot satisfy the unification condition (\ref{eq:normal}) exactly. 
This means the complete gauge unification requires slight corrections 
to the culculations of the gauge couplings, which can originate from 
$\alpha'$-corrections, KK-modes and so on. 
We have another alternative; the complete unification is not necessary to 
enjoy the benefit of nonuniversal gaugino masses, but also in this case, 
the three gauge couplings are almost unified within low-scale SUSY breaking scenarios. 
Therefore, the irrational number $x=(8\pm\sqrt{59})/30$ must be 
pretty close to a rational number 
which is ideal for theoretical consistency. 
In the following, we adopt $x=(8+\sqrt{59})/30$ even though it cannot be accepted 
since it is sufficient for our purpose to demonstrate 
the nonuniversarity of gaugino masses.

\begin{figure}[t]
\begin{center}
\includegraphics[width=0.5\linewidth]{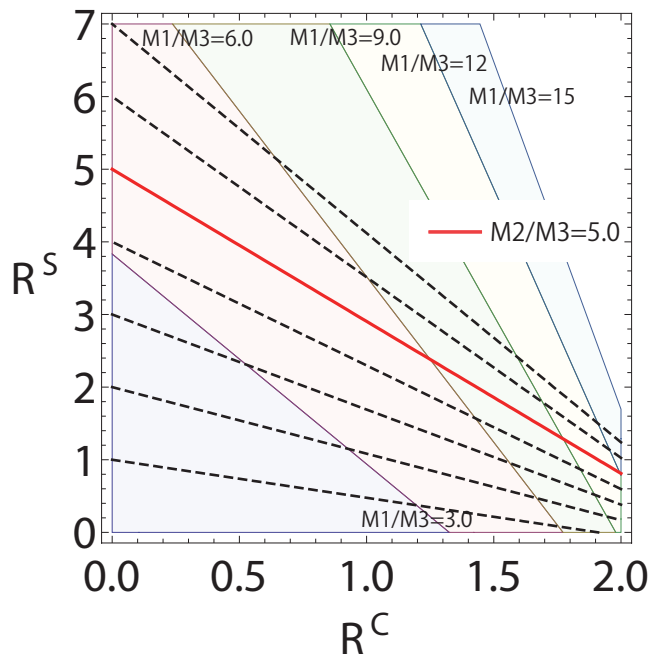}
\caption{The seven transversal lines represent the ratio of the wino mass 
to the gluino mass, corresponding to $M_2/M_3 = 1,2,\ldots, 7$ from the bottom line. 
The red solid line, which is one of the seven contours, corresponds to 
$M_2/M_3=5.0$. 
The colored shades show the other ratios, $M_1/M_3$. 
We obtain the ratio $0\le M_1/M_3\le 3$ in the purple region, 
and $3\le M_1/M_3\le 6$, $6\le M_1/M_3\le 9$, $9\le M_1/M_3\le 12$ 
and $12\le M_1/M_3\le 15$ 
are obtained in the red, green, yellow and cyan regions, respectively.  }
\label{fig:massratio}
\end{center}
\end{figure}

We estimate the ratios $M_1/M_3$ and $M_1/M_3$ on $(R_S, R_C)$-plane 
as shown in Fig.~\ref{fig:massratio}. 
In this figure, the mass ratio of the wino and gluino masses are represented by 
six black dashed lines and a red solid line. 
These seven contours correspond to $M_2/M_3=1.0,2.0,\ldots,7.0$ from the bottom, 
and the red solid line indicates the most favored one $M_2/M_3=5.0$. 
Five colored regions of this plane represent the mass ratio of the bino and the gluino. 
Each of the purple, red, green, yellow and cyan regions 
corresponds to $0\le M_1/M_3\le 3.0$, $3.0\le M_1/M_3\le 6.0$, 
$6.0\le M_1/M_3\le 9.0$, $9.0\le M_1/M_3\le 12.0$ and $12\le M_1/M_3\le 15$, respectively. 

Compared with to the favored mass ratios (\ref{eq:ratio1}), 
the bound $M_1/M_3\ge-3$ is always satisfied in all the parameter space, 
and the other $3.0\le M_2/M_3 \le 5.5$ is also realized in the wide region of the parameter space. 
In particular, the very natural region $5.2\le M_2/M_3 \le 5.5$ can be available. 
We see that the mechanism to relax the fine-tunning 
by the nonuniversal gaugino masses will work correctly in our model. 

\section{Conclusions and Discussions}
We have constructed a model of nonuniversal gaugino masses 
on the basis of a mixture of 6D $U(4)$ SYM theory and 
10D $U(4)$ SYM theory compactified on three 2D-tori with magnetic fluxes, 
motivated by stable D-brane systems. 
In higher-dimensional SYM systems, 
the toroidal compactifications with magnetic fluxes are able to generate 
the flavor structure of the SM, and SUSY preserving configurations of magnetic fluxes 
lead to MSSM-like models. 
In such models derived from single SYM theories, 
the gauge kinetic functions of the three SM gauge fields 
are given by a single modulus universally\footnote{
This is valid in SYM theories, but is nontrivial in string models, such as, 
D-brane models. 
Corrected gauge kinetic functions might be able to be calculated 
with DBI actions and so on in magnetized toroidal compactifications, 
but it is beyond the scope of this paper. }, 
and their couplings have the exactly same form. 
The gaugino masses must then degenerate at a certain high-energy scale. 
However, in 4D effective actions derived from two SYM theories defined in 
different dimensional spacetime, 
the moduli couplings to the SM gauge fields are involved by two moduli fields, 
and the three gauge kinetic functions are distinguishable, 
which generate the non-universal gaugino masses.

Our model leads to nonuniversal gauge kinetic functions containing 
two moduli fields, where we have studied the gauge coupling constants 
and the gaugino masses. 
Since we have found that the unification of gauge coupling constants 
can be roughly realized as is in the MSSM, 
 the favored ratios of gaugino masses estimated in the MSSM~\cite{Abe:2012xm} are 
also being reliable in our model. 
This is supported by another calculation given in the Appendix. 
The numerical analysis of gaugino masses has shown that 
our model can yield the gaugino masses with the ideal ratios. 
Thus, we have proposed a concrete model of nonuniversal gaugino masses 
in a top-down approach on the basis of a magnetized toroidal compactification of SYM system, 
where the mechanism to enhance the Higgs boson mass and relax the fine-tunning will work.

The conventional parameter space of the MSSM 
was certainly well motivated to study primarily. 
However, we see that 
the conventional parameter space gives 
mass spectra mostly accompanied by any of troublesome today 
in the light of the latest experimental and observational data, 
such as, the fine-tunning of $\mu$-parameter and lack of candidates for the dark matter. 
In this situation, 
unconventional parameter space of the MSSM has been getting attractive. 
Indeed, nonuniversal gaugino mass is one of them 
and the attractive prospect has been found. 
We should then remark on how to realize such an unconventional spectrum 
desirable from the phenomenological point of view, 
strictly, how to obtain unconventional boundary conditions for 
parameters of the MSSM at a certain high scale. 
A possibility to realize such values of parameter in theories for a high-energy physics 
should be taken into account more seriously. 
Such discussion will lead to restrictions on some of unconventional parameter space, 
or conversely, it is expected that 
we can infer theories describing a high-energy physics, which we cannot directly reach today, 
on the basis of the obtained experimental data.

\subsection*{Acknowledgement}
K.S. was supported by a Grant-in-Aid for JSPS Fellows 
No.~25$\cdot$4968 from the Ministry of Education, Culture, Sports, Science 
and Technology in Japan.

\appendix
\section{RGE effects on the gauge coupling constants} 
This appendix calculates the RG effects on gauge coupling constants in more detail. 
In our model, the coupling constants of $SU(3)_C$ and $SU(2)_L$ gauge symmetries 
are given by the 6D and 10D SYM couplings, respectively. 
A linear combination of the two gauge kinetic functions 
can be identified with that of the $U(1)Y$ gauge symmetry. 
In the combination, there remains the parameter $x$ with which 
we can control the gauge coupling constant and gaugino mass simultaneously. 
This should be fixed to satisfy relations 
between the gauge couplings $\tilde g_6=g_3$, $\tilde g_{10}=g_2$ and $g_1$. 
Indeed, we adopted a certain value of $x$ to obtain the gauge coupling unification in Sec.~5. 
In other words, we can always choose this parameter to approximately realize 
the experimental value of $U(1)_Y$ coupling constant. 
Thus, we are allowed to concentrate on the other two gauge coupling constants here. 

We solve the RGEs for the gauge coupling constants of $SU(3)_C$ and $SU(2)_L$. 
The two experimental values evolve in accordance with the SM RGEs 
from the electroweak scale $\Lambda_{\rm EW}$ 
to a typical SUSY breaking scale $\Lambda_{\rm SUSY}$. 
From $\Lambda_{\rm SUSY}$ scale to the compactification scale defined as 
$M_C = 1/\sqrt{\mathcal A^{(1)}}$, 
we can calculate the running of gauge couplings by using 
the RGEs within the MSSM. 
Lastly, the obtained values at $M_C$ scale will further evolve to the string scale 
within $U(4)$ YM theories, 
and these will be consistent with theoretical calculation shown in 
Eq.~(\ref{eq:gauge}). 
For $SU(3)_C$ gauge, these are summarized as 
\begin{equation}
4\pi\frac{e^{-\phi}}{(2\pi)^3\alpha'} \mathcal A^{(1)} = 
-\frac {b_5}{2\pi}(t_{\rm st} -t_{\rm C})
-\frac {\tilde b_3}{2\pi}(t_{\rm C} -t_{\rm SUSY})
-\frac {b_3}{2\pi}(t_{\rm SUSY} -t_{\rm EW})
+\alpha^{-1}_{3\exp},\label{eq:gg5}
\end{equation}
where the coefficient of beta functions,$b_3$ and $\tilde b_3$, 
are given by $-7$ (SM) and $-3$ (MSSM). 
The coefficient $b_5=-4$ is related to the $U(4)_C$ SYM theory which contains 
an adjoint fields and two bifundamental representations\footnote{
This calculation does not contain the relevant KK modes. 
Their contributions would be negligible because 
the compactification scale and the string scale are almost 
equal in any cases.  }. 
The experimental value of gauge coupling is defined by 
$\alpha^{-1}_{3\exp}\equiv 4\pi/g_{3,\exp}^2$. 
Each energy scale $t$ is defined as follows, 
\begin{eqnarray*}
t_{\rm st} &\equiv& \log (M_{st}/\Lambda_0)\\
t_{\rm C} &\equiv& \log (M_{C}/\Lambda_0)\\
t_{\rm SUSY} &\equiv& \log (M_{\rm SUSY}/\Lambda_0)\\
t_{\rm EW} &\equiv& \log (M_{Z}/\Lambda_0), 
\end{eqnarray*}
where $\Lambda_0$ is an input scale and 
$M_{Z}$ represents the mass of $Z$ boson.

For $SU(2)_L$ coupling constant, 
the similar relation between, the experimental values of $SU(2)_L$ coupling constant and 
theoretical representation given at the string scale, is given by 
\begin{equation}
4\pi\frac{e^{-\phi}}{(2\pi)^7{\alpha'}^3} \mathcal A^{(1)}\mathcal A^{(2)}\mathcal A^{(3)} = 
-\frac {b_9}{2\pi}(t_{\rm st} -t_{\rm C})
-\frac {\tilde b_2}{2\pi}(t_{\rm C} -t_{\rm SUSY})
-\frac {b_2}{2\pi}(t_{\rm SUSY} -t_{\rm EW})
+\alpha^{-1}_{2\exp},\label{eq:gg9}
\end{equation}
where the coefficients are given as $b_9=4$, $\tilde b_2=1$ and 
$b_2=-19/64$. 

Eqs.~(\ref{eq:gg5}) and (\ref{eq:gg9}) can be more simplified 
by using the 4D Planck mass (\ref{eq:mplanck}), the SUSY condition 
$\mathcal A^{(1)}/\mathcal A^{(2)}=\mathcal A^{(1)}/\mathcal A^{(3)}=3$, 
and typical values of $M_{\rm SUSY}=10^3 {\rm GeV}$ and $M_{\rm Z}=10^2 {\rm GeV}$, 
and then, they will be simultaneous equations for $M_{\rm st }$ and $M_C$. 
As the result, we find 
\begin{eqnarray*}
M_{\rm st } &=& 2.78\times 10^{18} {\rm GeV}\\
M_{\rm C } &=& 2.65\times 10^{17} {\rm GeV}. 
\end{eqnarray*}
This calculation of gauge couplings leads to the almost same result as 
one obtained in assumption of the rough gauge coupling unification 
shown in Eqs.~(\ref{eq:mst}) and (\ref{eq:mcc}). 
This means that deviation from the MSSM in terms of the running of gauge 
couplings is small enough for 
the ideal ratios of gaugino masses (\ref{eq:ratio1}) 
estimated in the MSSM 
to be applicable to our model directly.


\begin{thebibliography}{99}

\bibitem{Aad:2012tfa}
  G.~Aad {\it et al.}  [ATLAS Collaboration],
  Phys.\ Lett.\ B {\bf 716} (2012) 1
  [arXiv:1207.7214 [hep-ex]].

\bibitem{Chatrchyan:2012ufa}
  S.~Chatrchyan {\it et al.}  [CMS Collaboration],
  Phys.\ Lett.\ B {\bf 716} (2012) 30
  [arXiv:1207.7235 [hep-ex]].

\bibitem{Abe:2007kf}
  H.~Abe, T.~Kobayashi and Y.~Omura,
  Phys.\ Rev.\ D {\bf 76} (2007) 015002
  [hep-ph/0703044 [HEP-PH]]

\bibitem{Abe:2012xm}
  H.~Abe, J.~Kawamura and H.~Otsuka,
  PTEP {\bf 2013} (2013) 013B02
  [arXiv:1208.5328 [hep-ph]].

\bibitem{Antusch:2012gv}
  S.~Antusch, L.~Calibbi, V.~Maurer, M.~Monaco and M.~Spinrath,
  JHEP {\bf 1301} (2013) 187
  [arXiv:1207.7236].

\bibitem{Bachas:1995ik}
  C.~Bachas,
  hep-th/9503030.

\bibitem{Cremades:2004wa}
  D.~Cremades, L.~E.~Ibanez and F.~Marchesano,
  JHEP {\bf 0405} (2004) 079  [hep-th/0404229].



\bibitem{Abe:2012fj}
  H.~Abe, T.~Kobayashi, H.~Ohki, A.~Oikawa and K.~Sumita,
  Nucl.\ Phys.\ B {\bf 870} (2013) 30
  [arXiv:1211.4317 [hep-ph]].

\bibitem{Abe:2014soa}
  H.~Abe, J.~Kawamura and K.~Sumita,
  Nucl.\ Phys.\ B {\bf 888} (2014) 194
  [arXiv:1405.3754 [hep-ph]].

\bibitem{Abe:2012ya} 
  H.~Abe, T.~Kobayashi, H.~Ohki and K.~Sumita,
  Nucl.\ Phys.\ B {\bf 863}, 1 (2012)
  [arXiv:1204.5327 [hep-th]].

\bibitem{Abe:2015jqa}
  H.~Abe, T.~Horie and K.~Sumita,
  arXiv:1507.02425 [hep-th].



\bibitem{Barbieri:1987fn}
  R.~Barbieri and G.~F.~Giudice,
  Nucl.\ Phys.\ B {\bf 306} (1988) 63.


\bibitem{Randall:1998uk}
  L.~Randall and R.~Sundrum,
  Nucl.\ Phys.\ B {\bf 557} (1999) 79
  [hep-th/9810155]. 

\bibitem{Giudice:1998xp}
  G.~F.~Giudice, M.~A.~Luty, H.~Murayama and R.~Rattazzi,
  JHEP {\bf 9812} (1998) 027
  [hep-ph/9810442].

\bibitem{Dine:1981gu}
  M.~Dine and W.~Fischler,
  Phys.\ Lett.\ B {\bf 110} (1982) 227; 
  C.~R.~Nappi and B.~A.~Ovrut,
  Phys.\ Lett.\ B {\bf 113} (1982) 175; 
  L.~Alvarez-Gaume, M.~Claudson and M.~B.~Wise,
  Nucl.\ Phys.\ B {\bf 207} (1982) 96.

\bibitem{Abe:2009dr}
  H.~Abe, K.~S.~Choi, T.~Kobayashi and H.~Ohki,
  JHEP {\bf 0906} (2009) 080
  [arXiv:0903.3800 [hep-th]].


\bibitem{Marcus:1983wb}
  N.~Marcus, A.~Sagnotti and W.~Siegel,
  Nucl.\ Phys.\ B {\bf 224} (1983) 159.

\bibitem{ArkaniHamed:2001tb}
  N.~Arkani-Hamed, T.~Gregoire and J.~G.~Wacker,
  JHEP {\bf 0203} (2002) 055
  [hep-th/0101233].

\bibitem{Ibanez:2012zz}
  L.~E.~Ibanez and A.~M.~Uranga,
  Cambridge, UK: Univ. Pr. (2012) 673 p

\bibitem{Abe:2008fi}
  H.~Abe, T.~Kobayashi and H.~Ohki,
  JHEP {\bf 0809} (2008) 043
  [arXiv:0806.4748 [hep-th]].

\bibitem{Abe:2008sx}
  H.~Abe, K.~S.~Choi, T.~Kobayashi and H.~Ohki,
  Nucl.\ Phys.\ B {\bf 814} (2009) 265
  [arXiv:0812.3534 [hep-th]].

\bibitem{Abe:2015yva}
  T.~h.~Abe, Y.~Fujimoto, T.~Kobayashi, T.~Miura, K.~Nishiwaki, M.~Sakamoto and Y.~Tatsuta,
  Nucl.\ Phys.\ B {\bf 894} (2015) 374
  [arXiv:1501.02787 [hep-ph]].

\bibitem{Choi:2005uz}
  K.~Choi, K.~S.~Jeong and K.~-i.~Okumura,
  JHEP {\bf 0509} (2005) 039
  [hep-ph/0504037]. 

\bibitem{Endo:2005uy}
  M.~Endo, M.~Yamaguchi and K.~Yoshioka,
  Phys.\ Rev.\ D {\bf 72} (2005) 015004
  [hep-ph/0504036].




\end{thebibliography}
\end{document}